\documentclass[aps,prb,twocolumn,amsmath,amssymb]{revtex4-1}
\usepackage{graphicx}% Include figure files

\newcommand{\bmo}{BaMn$_2$O$_3$}

\begin{document}
\title{
Magnetostrictive N{\'e}el ordering of the spin-5/2 ladder compound
\bmo : \\ distortion-induced lifting of geometrical frustration }

\author{M.~Valldor$^1$, O.~Heyer$^1$, A.~C.~Komarek$^1$, A.~Senyshyn$^2$, M.~Braden$^1$,  and T.~Lorenz$^1$}
\affiliation{$^1$II.~Physikalisches Institut, Universit\"{a}t zu
K\"{o}ln, Z\"{u}lpicher Str.~77, 50937 K\"{o}ln, Germany\\
$^2$Technische Universit\"{a}t Darmstadt, Material und
Geowissenschaften, Petersenstr.~23, 64287 Darmstadt, Germany
and Technische Universit\"{a}t M\"{u}nchen, FRM-II,
Lichtenbergstr.~1, 85747 Garching, Germany}

\date{\today}

\begin{abstract}

The crystal structure and the magnetism of  BaMn$_2$O$_3$ have
been studied by thermodynamic and by diffraction techniques using
large single crystals and powders. BaMn$_2$O$_3$ is a realization of a $S =
5/2$ spin ladder as the magnetic interaction is dominant along
180$^\circ$ Mn-O-Mn bonds forming the legs and the rungs of a
ladder. The temperature dependence of the magnetic susceptibility
exhibits well-defined maxima for all directions proving the
low-dimensional magnetic character in \bmo . The susceptibility
and powder neutron diffraction data, however, show that \bmo \
exhibits a transition to antiferromagnetic order at 184~K, in
spite of a full frustration of the nearest-neighbor inter-ladder
coupling in the orthorhombic high-temperature phase. This
frustration is lifted by a remarkably strong monoclinic distortion
which accompanies the magnetic transition.

\end{abstract}


%75.10.Jm 	Quantized spin models, including quantum spin frustration 
%75.40.Cx 	Static properties (order parameter, static susceptibility, heat capacities, critical exponents, etc.) 
%75.80.+q 	Magnetomechanical effects, magnetostriction

\maketitle

\section{Introduction}

Exotic magnetic ground states are expected and have been observed
in a special group of layered metal oxides containing spin
ladders\cite{hiroi91,dagotto92,dagotto96,white02,carr02,vuletic06}. 
In the cuprates, these quasi-one
dimensional magnetic units of $S = 1/2$ ($3d^9$) Cu$^{2+}$ ions
belong to the most well studied spin liquids and do
not evolve long-range magnetic order at finite temperatures. This
mainly arises from a very effective magnetic decoupling of
neighboring spin ladders due to a geometrical frustration of the
magnetic spin interactions present between the
ladders. The strongest antiferromagnetic (AFM)
spin-spin couplings $J_1$ and $J_2$ are expected, respectively,
along the rungs and legs of the spin ladders due to almost perfect
super-exchange conditions\cite{goodenough55,kanamori59}, which are
responsible for interesting spin dynamics\cite{eccleston96}. Small
anisotropies in J within the ladders could also result in spin
dimerization\cite{martin-delgado98}. Furthermore, due to the low
dimensionality of such spin-ladder materials they can act as model systems and
connect theory with experiment in e.g. spectroscopy\cite{Windt01,Klanjsek08,Ruegg08cm,Thielemann09PRL},
thermal conductivity\cite{Sologubenko00,Hess01}, and
thermodynamics\cite{Lorenz08,Anfuso08}. The spin-ladder structural
features are still rare in nature, but are present in the metal
oxide \bmo \cite{sander79}. The high spin moment $S = 5/2$ of
Mn$^{2+}$ ($3d^5$) represents a very different situation as for
the cuprates, but theoretical work on the $S = 5/2$ model systems
predicts dimerization\cite{tangoulis07}. Here, we present a
combined study of magnetism, x-ray and neutron diffraction, and
thermodynamic properties on \bmo.

\begin{figure}
\includegraphics[width= .8\linewidth]{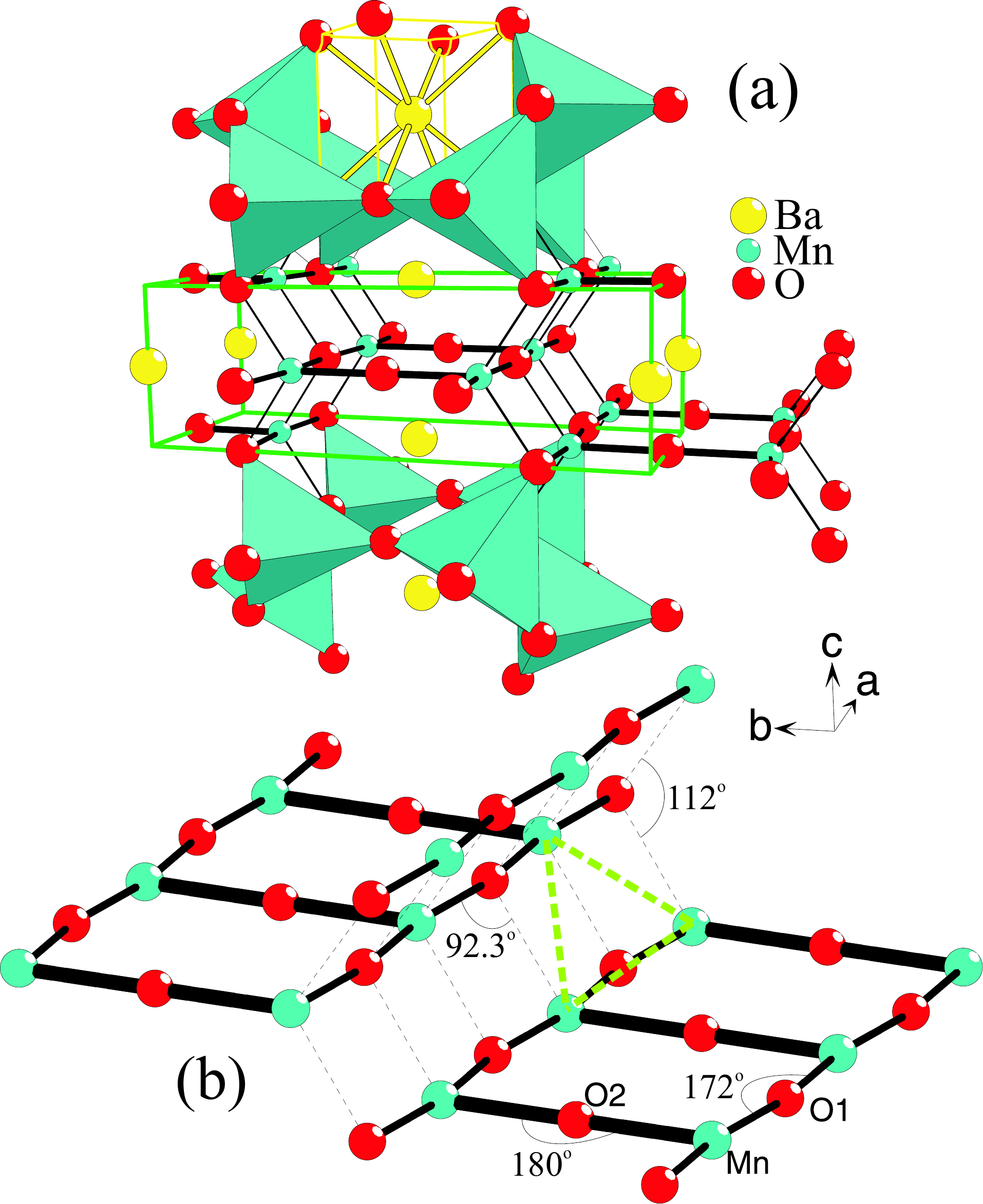}
\caption{\label{fig1} (Color online) (a) Room temperature structure of \bmo. The
red balls at the corners of the Mn-containing polyhedra are oxygen
atoms. The unit cell is indicated in green. The eightfold Ba
coordination is indicated by yellow lines. (b) A selected part of
the structure showing the Mn-O substructure with Mn--O--Mn bond
angles. The dashed green lines indicate the geometrical
frustration, as discussed in the text.}
\end{figure}

The room-temperature crystal structure of \bmo\ has been described
by Sander and M{\"u}ller-Buschbaum\cite{sander79} and is shown in
Fig.~\ref{fig1}~(a). Similar to the hexagonal $R$MnO$_3$ ($R =
$~Ho--Lu, Sc, and Y)\cite{zhou06}, Mn has a five-fold
coordination, which forms an almost perfect trigonal bipyramid.
However, the relatively lower oxygen content in \bmo, i.e.\ the
lower oxidation state of Mn, forces the MnO$_5$ polyhedra to also
share edges instead of only corners. This condensation is known as
"shear-planes" in tungstenates\cite{kihlborg68} and causes the
structure to evolve double layers of edge-sharing MnO$_5$
polyhedra, perpendicular to the $b$ axis, with corner sharing as
the only connection between these double layers. The Ba$^{2+}$
ions fill the space between the layers and coordinate eight oxygen
ions. The Mn--O--Mn super-exchange, however, is not governed by
the double layers but by the bond angles and by the bond
directions, shown in Fig.~\ref{fig1}~(b), which amount to
180$^\circ$ [010], 172$^\circ$ [100], 112$^\circ$ [001], and
92.3$^\circ$, which should ensure strong magnetic coupling along
the rungs ($J_1$) and the legs ($J_2$) of each ladder, but weaker
interactions between them. The Mn--to--Mn interactions from one
ladder to the nearest-neighboring ones constitute a complete
geometric frustration. By symmetry the interactions are equally
strong to the two next-nearest Mn$^{2+}$ ions in the
nearest-neighboring ladder (thick dashed lines in
Fig.~\ref{fig1}b). This results in a geometric frustration and in
an effective decoupling of the neighboring ladders, and quasi 1D
AFM interactions along the [100] axis are expected to dominate the
magnetic correlations of \bmo .

\section{Experimental}
\label{sec:exp}

A single crystal of centimeter size was synthesized using a
floating-zone image furnace (FZ-T-10000-H-VI-VP) in a
flowing gas mixture of N$_2$/H$_2$(5\%) (0.25~dm$^3$/min). Both
feeding bar and seed were pressed from stoichiometric amounts of
BaCO$_3$ (Strem Chem. 99.9\%) and MnO (Aldrich 99+\%).
The issue of BaO loss due to a transport
reaction with H$_2$, as was mentioned in the first paper on this
compound\cite{sander79}, could to some extent be avoided by growing the
crystal at a relatively high rate of 10~mm/h. Indeed, the obtained
crystal was covered by a thin layer of lime green crystals (MnO),
indicating a small loss of BaO, but this top layer could be removed
mechanically. Such a pure \bmo\ crystal appears dark, almost black, but thinner parts reveal that the
compound is transparent and dark forest-green. The pale yellow
color observed in Ref.~\onlinecite{sander79} can be
seen if the crystals are further crushed, reaching micrometer
size. The compound decomposes slowly in air, but relatively fast
in water. Both reactions cause the crystal surface to become
brown, probably due to formation of Ramsdellite ($\alpha$-MnO$_2$)
and barium hydroxides and carbonates.

Elemental analysis was performed in a scanning electron microscope
SEM515 (Philips) equipped with an EDX unit at 20~kV acceleration
voltage. The average metal-to-metal composition was determined to
be Ba$_{1.002(8)}$Mn$_{1.998(8)}$ from 10 EDX analyses at
different spots on a centimeter-sized single-crystal slice; this
very well agrees with the expected metal stoichiometry. The purity
of the sample was checked by powder X-ray diffraction and no
impurity phases could be detected meaning that the sample is at
least 95\% pure.

The X-ray powder diffraction data were collected at 300~K using a Cr
$K_{\alpha_{1,2}+\beta}$ X-ray tube ($\lambda = 2.28973$, 2.29365,
and 2.08090~\AA) as source. The intensities were
collected in reflection (Bragg-Brentano) geometry. Single crystal X-ray measurements 
were performed down to 100~K with a Bruker X8 APEX
(Mo$K_{\alpha_{1,2}}$, $\lambda = 0.7093$ and 0.71359~\AA); the
cooling was accomplished by using a flow of dry N$_2$ gas
(Cryoflex). Neutron diffraction data was collected at
SPODI, FRM II (Munich, Germany), with a constant wavelength of
2.537 \AA (Ge [331] monochromator) at temperatures between 3 and
300~K. A vanadium cup was used as sample holder for the powdered
single crystal sample.

Magnetic and specific-heat measurements were performed
in a commercial Physical-Properties-Measurement-System (PPMS, Quantum
Design Inc.) in the temperature range 2-400~K in magnetic fields
up to 14 T. The magnetization was measured by the vibrating-sample
technique while either the magnetic field or the temperature is
continuously varied. For the specific heat, a relaxation-time
method is used and the data points are typically obtained
step-wise after stabilizing certain temperatures. In addition, we
also used a quasi-continuous modification of this relaxation-time
method (see Ref.~\onlinecite{lashley03}) in the temperature range around
the first-order phase transition. Thermal
expansion was investigated on a home-built high-resolution
capacitance dilatometer.

To ensure that the sample did not deteriorate in air, the crystal
was only handled in an Ar-filled glove box and sealed in a shrink
hose before performing the magnetic measurements. For all the
other methods the crystals were rapidly transferred in air from
the Ar-filled glove box to the respective measurement setups,
which work either in He atmosphere, dry N$_2$ gas or under vacuum
conditions. In none of the measurements on \bmo\ single crystals
we could observe indications for the presence of a sizeable amount
of impurity phases. This was different for the neutron powder
diffraction data which revealed that the studied sample contains
about 3~\% of MnO. This impurity phase in the neutron data
probably originates from the fact that the exposure to air is more
severe for a powdered sample with a much larger surface-to-volume
ratio than a single crystal.

\section{Thermodynamic properties}

\begin{figure}
\includegraphics[width= .9\linewidth]{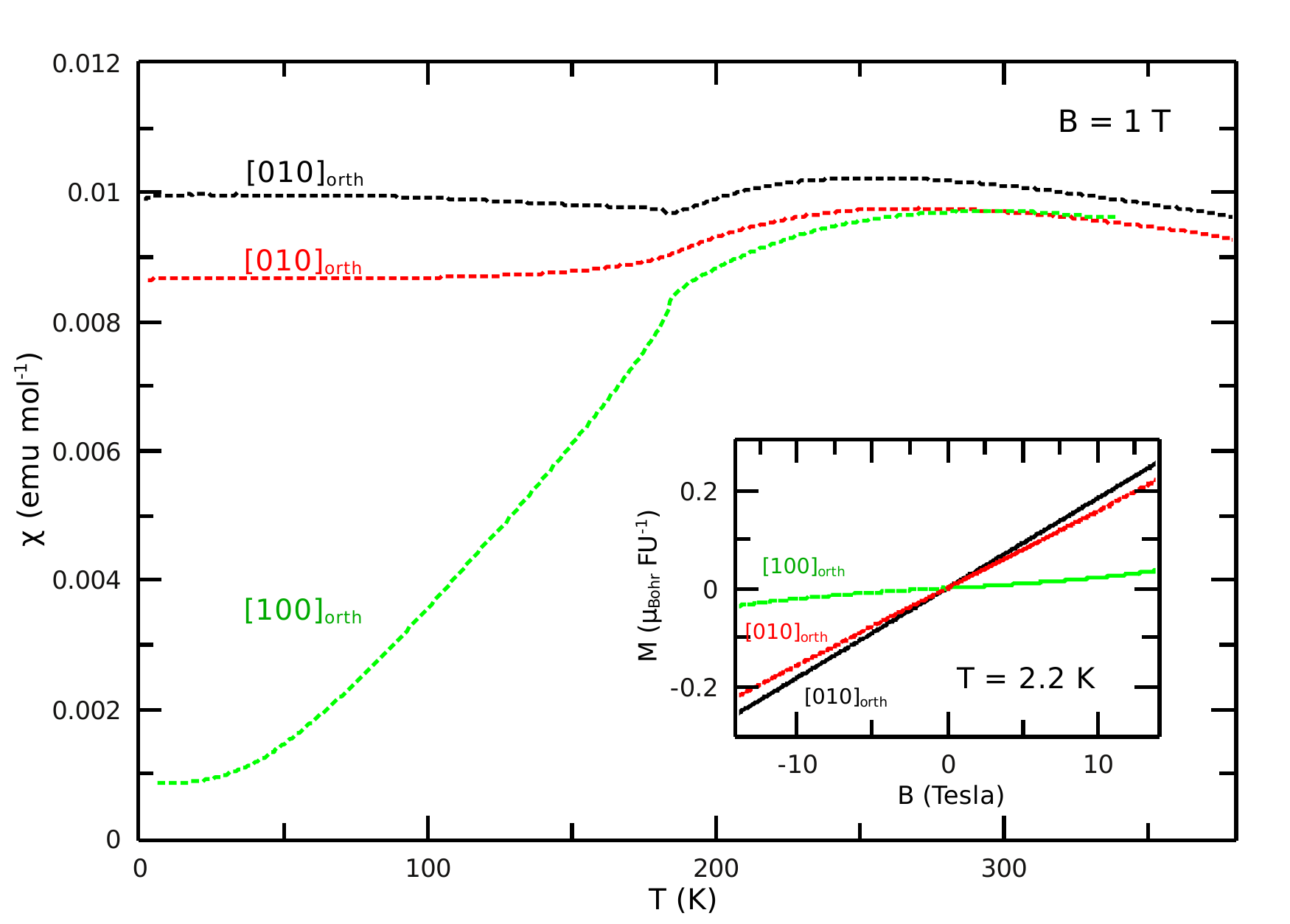}
\caption{\label{fig4} (Color online) Magnetic susceptibility as a function of
temperature measured in magnetic fields of 1~T applied along all
three orthorhombic crystallographic directions of \bmo. The inset
displays magnetization curves measured at 2.2 K as a function of
the magnetic field up to 14~T applied along the orthorhombic
axes.}
\end{figure}

Fig.~\ref{fig4} summarizes the magnetic measurements. Between
about 200 and 400~K, the magnetic susceptibility $\chi$ shows a
weak anisotropy with respect to the direction of the applied
magnetic field. The data of Fig.~\ref{fig4} have been obtained
after cooling the sample in zero magnetic field. The corresponding
data from field-cooling experiments (not shown) perfectly
superimpose these data for all three field directions, which rules
out the existence of some type of ferromagnetic domains or a
spin-glass behavior in \bmo. For all three field directions, the
temperature dependencies $\chi_i(T)$ strongly deviate from a
simple Curie-Weiss behavior. Instead the $\chi_i(T)$ show broad
maxima which, depending on the field direction $i$, are located
between about 250 and 290~K. These broad maxima are typical for
low-dimensional magnets and signal the continuous increase of
magnetic correlations with decreasing temperature. The
susceptibility data unambiguously proves the low-dimensional
nature of magnetic correlations in \bmo . At $\simeq 184$~K clear
anomalies occur in all three $\chi_i(T)$: for a field along the
orthorhombic $a$ axis, $\chi_a$ strongly decreases and levels off
at a very small value at the lowest temperature, while for a field
along the orthorhombic $b$ ($c$) axis $\chi_i$ slightly increases
(decreases) below 184~K and finally remains essentially constant
below about 100~K. This anisotropic behavior is a clear indication
of an antiferromagnetic ordering, where, below the N{\'e}el
temperature $T_N\simeq 184$~K, the spins spontaneously align
predominantly (anti)parallel to the orthorhombic $a$ axis.

The low-temperature magnetization measured as functions of
magnetic fields up to 14~T applied along these three field directions
are shown in the inset of Fig.~\ref{fig4}. Obviously, there are
neither indications of metamagnetic transitions nor of a magnetic
hysteresis. Instead, these data show that the low-field anisotropy
observed in the temperature-dependent measurements at 1~T remains
essentially preserved up to a field of 14~T. At first glance, this is a surprising
observation, because for the $3d^5$ configuration of Mn$^{2+}$ the
orbital contribution to the magnetic moment is typically small and
consequently the magnetic anisotropy is expected to be weak. In
other words, Mn$^{2+}$ systems are expected to represent rather
well the spin-5/2-Heisenberg system. In this case, however, only a
weak magnetic anisotropy is expected and comparatively low magnetic
fields along the easy $a$ axis should be sufficient to induce a
so-called spin-flop transition, where the spin orientation changes
from being (anti)parallel to the $a$ axis to a direction within the
plane perpendicular to $a$.

A rough estimate of the effective intra-ladder couplings can be
obtained by comparing the positions of the susceptibility maxima
of \bmo\ to Monte Carlo simulations of $S=5/2$
ladders\cite{tangoulis07}. According to these simulations the
susceptibility maximum is expected to occur at $T\simeq 6.69 J$
for a homogeneous spin ladder with equal rung and leg coupling
$J_1=J_2=J$. Thus, our data suggest an intra-ladder coupling
$J\simeq 40$~K. Due to the large spin value, this coupling
corresponds to a very large saturation field
$H_{sat}=2Sk_B(2J_1+J_2)/g\mu_B\simeq 450$~T. In view of this
large saturation field, the absence of a spin-flop transition in
the studied field range appears less surprising, because from
the present data one can only conclude that the spin-flop field
$H_{SF}$ is larger than 14~T, i.e.\ $H_{SF}/H_{sat} > 3$~\%.
Within a mean-field treatment\cite{dejongh74}, this can be
explained already by a very weak anisotropy field $H_{A}/H_{sat} >
10^{-3}$. A closer inspection of the data in the inset of
Fig.~\ref{fig4} reveals a weak increase of slope in the high-field
region of the magnetization curve for a field applied along $a$.
Thus, one may speculate that a spin-flop transition could occur in
the field range slightly above our maximum field. In order to
clarify this issue, measurements to higher fields are necessary.

\begin{figure}
\includegraphics[width= 1.\linewidth]{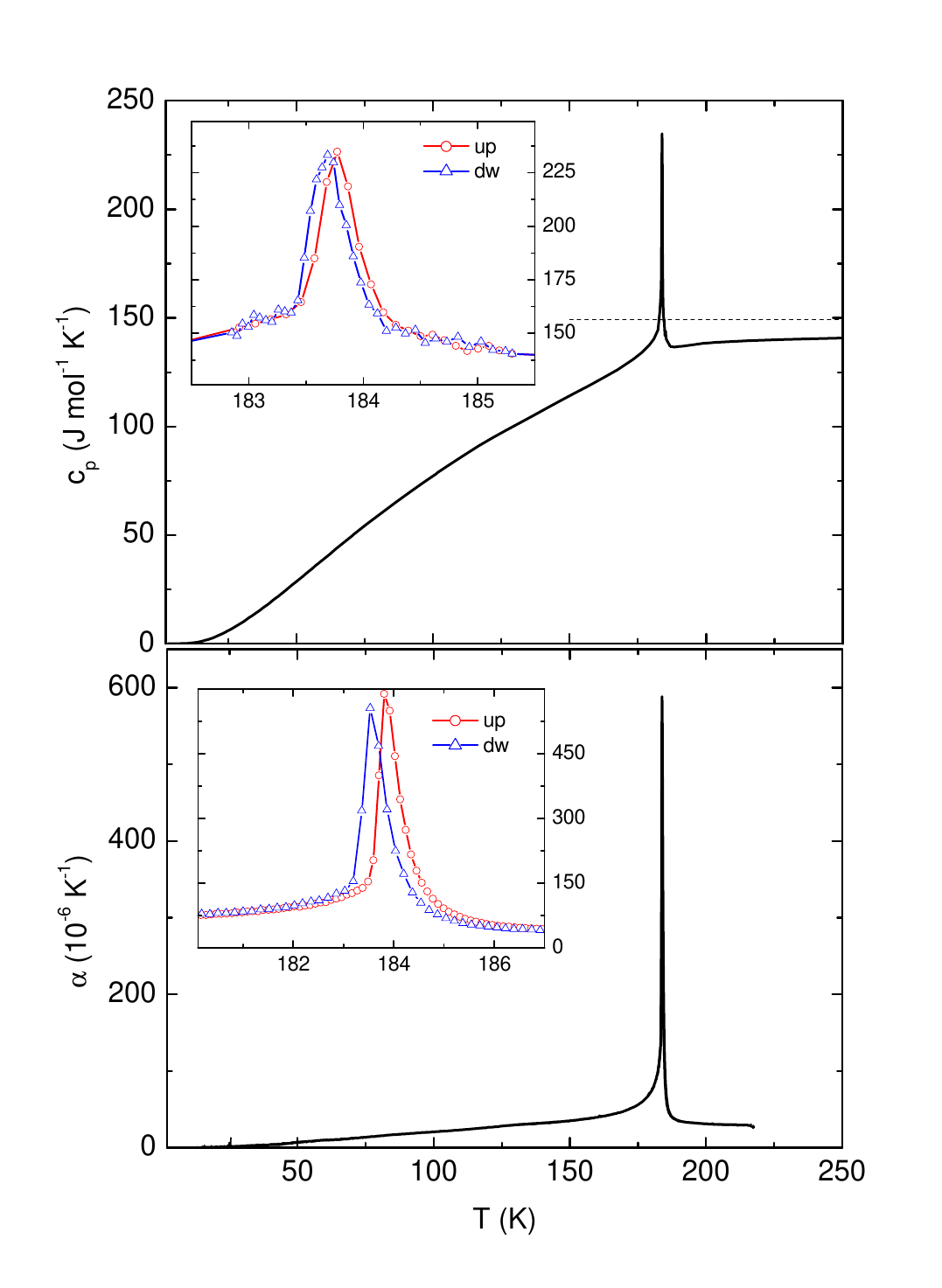}
\caption{\label{fig3} (Color online)  Specific heat (top) and linear thermal expansion (bottom)
of \bmo\ in zero magnetic field. Thermal expansion was measured along  
[001]$_{orth}$ direction. The dashed line in the upper panel marks the  
Dulong-Petit limit. In both panels, the inset is a blow-up of the respective 
heating (up) and cooling (dw) curves in the temperature range
close to the N{\'e}el ordering.}
\end{figure}

\begin{table*}[t]
\caption{Single crystal X-ray data of \bmo\ at 150 and 100 K. Each
atomic site is presented as follows: Atom, occ., $x$, $y$, $z$,
$U_{iso}$, $U_{11}$, $U_{22}$, $U_{33}$, $U_{12}$, $U_{13}$,
$U_{23}$ with STDs in parentheses. (*adopted from the neutron
powder diffraction data at the same temperature).}
 \label{tab1}
%  \scriptsize
\vskip1mm
\parbox{16.5cm}{
\begin{ruledtabular}
\begin{tabular}{lll}
Temperature (K) & 150 & 100  \\
\hline
{\it 2a} & 	 
  Ba, 1.0, 0, 0, 0, 0.0069(1), 0.0074(1), & Ba,1.0, 0, 0, 0, 0.00512(9), 0.00569(9),  \\
 & 0.0069(1), 0.0071(1), 0, 0.004988(8), 0  &  0.00515(9), 0.00532(9), 0 0.00384(7), 0  \\
\hline
{\it 4g} & Mn, 1.0, 0, 0.31183(3), 0, 0.0068(2), 0.0068(2), & Mn, 1.0, 0, 0.31183(3), 0, 0.00432(8), 0.0056(1), \\
 & 0, 0.31183(3), 0, 0.00432(8), 0.0056(1), & 0.0050(1), 0.0059(1), 0, 0.0037(1), 0  \\
\hline 
{\it 2b} & O, 1.0, 0, 1/2, 0, 0.0092(9), 0.0098(9), & O, 1.0, 0, 1/2, 0, 0.0075(9), 0.0093(9),  \\
 & 0.0061(7), 0.0112(9), 0, 0.0065(8), 0 & 0.0043(7), 0.0094(8), 0, 0.0061(7), 0 \\
\hline 
{\it 4h} & O, 1.0, 0, 0.2036(2), 1/2, 0.0085(6), 0.0076(6),  & O, 1.0, 0, 0.2036(1), 1/2, 0.0069(6), 0.0063(5), \\
 & 0.0089(6), 0.0090(6), 0, 0.0053(5), 0 & 0.0071(5), 0.0074(5), 0, 0.0043(5), 0 \\
\hline 
Space group (Nr.)	& $C2/m$ (12)	& $C2/m$ (12) \\ \hline
 $a, b, c$ (\AA)*	& 5.62060(2), 10.96155(3), 3.53794(2)	& 5.61194(2), 10.95766(3), 3.53324(2)  \\ \hline
 $\alpha,\beta,\gamma$ ($^\circ$)* &	90, 128.8199(2), 90	& 90, 128.7261(1), 90  \\ \hline
% Parameters &	19	& 19 \\ \hline
 $S$ (obs, all)&	1.21, 1.21	& 1.16, 1.17   \\ \hline
% $S$ (all) &	1.21 &	1.17   \\ \hline
 $R$(obs, all) &	0.0189, 0.0189 &	0.0175, 0.0175   \\ \hline
% $R$(all)&	0.0189&	0.0175   \\ \hline
 $R_w$(obs, all)&	0.0546, 0.0547&	0.0518, 0.0518   \\ \hline
% $R_w$(all)  &	0.0547 &	0.0518   \\ \hline
 Refined str. factor &	F$^2$	& F$^2$   \\ \hline
  Z & 2 & 2  \\ \hline
 % Calc. $\rho$ &	5.7508 g/cm$^3$ &	5.7597 g/cm$^3$            \\ \hline
  Transm. min/max.	 & 0.0659/0.1291  &	0.0980/0.1714  \\ \hline
  Abs. coeff & 18.527 mm$^{-1}$	& 18.555 mm$^{-1}$  \\ \hline
  F(000) &	260	& 260  \\ \hline
 % $\theta$  range	3.71$^\circ$ -- 40.50$^\circ$ & 	3.71$^\circ$ -- 40.46$^\circ$   \\ \hline
 % $hkl$ range	& -10--9,-20--19,-6--5 & 	-10--9,-19--19,-6--5   \\ \hline
  Ref. total	& 561 &	563  \\ \hline
  Ref.($I>3\sigma$ )	& 557	& 561  \\ \hline
  Peak/hole &	0.87/-1.24 e/\AA$^3$	& 0.95/-1.25 e/\AA$^3$  \\ \hline
\end{tabular}
\end{ruledtabular}
} % wg. \parbox
\end{table*}

In the specific heat, shown in Fig.~\ref{fig3}a, a pronounced
anomaly appears close to 184~K, which signals a release in
entropy. The specific-heat peak is slightly asymmetric,
$\lambda$-like, which would infer that the phase transition is of
second order. However, measurements with increasing and decreasing
temperature reveal a small hysteresis of $\simeq 0.2$~K (inset of
Fig.~\ref{fig3}a), which identifies this transition as a weak
first-order one. In Fig.~\ref{fig3}b, we show the linear thermal
expansion measured along the [001] direction according to the
orthorhombic settings. Again, a strong anomaly is observed close
to 184~K, with a minor thermal hysteresis of about 0.2~K between
the curves measured on increasing and decreasing temperature, in
very good agreement with the results of the specific heat data.
The magnitude of the thermal-expansion anomaly at $T_N$ is huge,
which reveals that this magnetic ordering strongly couples to the
lattice. The strong magnetoelastic coupling is also the most
likely reason for the first-order nature of this phase transition.
It has been shown that an intrinsically second-order phase
transition can be driven to first order by a finite coupling to
the lattice\cite{kadanoff67,ito78}.

From the presented macroscopic data the behavior of \bmo\ may be
summarized as follows: \bmo\ contains 2-leg spin ladders with spin
5/2 and a rather strong intra-ladder coupling of the order of
$J\simeq 40$~K as inferred from the broad susceptibility maxima
around 270~K. Despite this large coupling, a three-dimensional
N\'eel order is suppressed because the effective inter-ladder
coupling is weak as a consequence of the geometrical arrangement
of neighboring ladders that causes strong magnetic frustration.
Nevertheless, \bmo \ undergoes an antiferromagnetic ordering transition at
T=184\ K. In order to allow for a three-dimensional N\'eel ordering, this
frustration can be lifted by a structural distortion, which then
naturally explains the strong magnetoelastic coupling and the
weakly first-order nature of the observed phase transition at
184~K. In order to resolve the microscopic details of these
structural changes and the magnetic structure we performed
single-crystal X-ray measurements down to 100~K as well as neutron
powder diffraction measurements on a crushed single crystal in the
temperature range from 3 to 300~K.

\section{Crystallographic and Magnetic Structure}

\begin{figure}
\includegraphics[width= .8\linewidth]{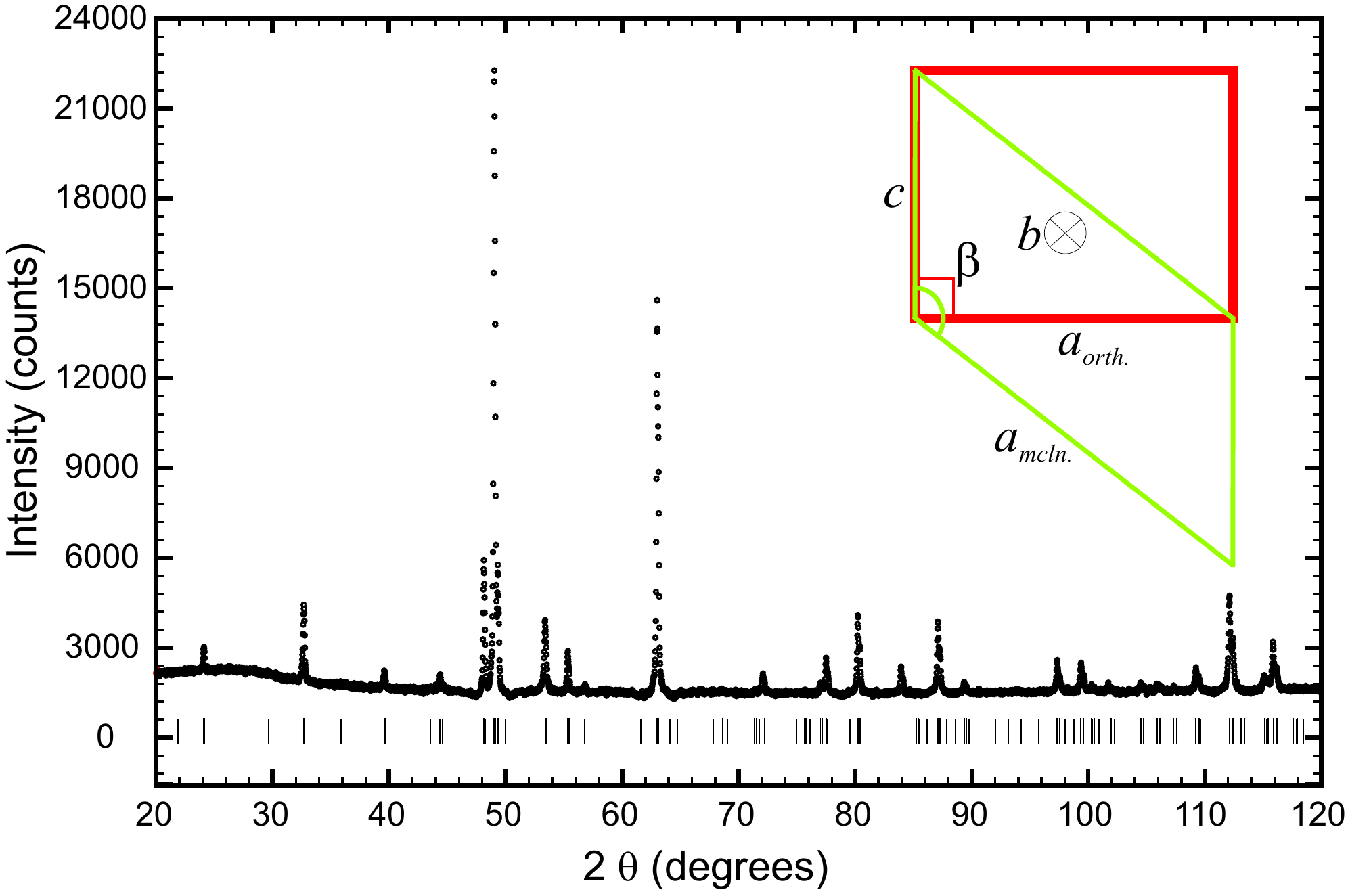}
\caption{\label{fig2} (Color online)  Room temperature X-ray powder diffraction
pattern of \bmo. The observed intensities are marked by open
circles and the expected Bragg positions by vertical lines. The
upper-right inset displays the orthorhombic and the monoclinic unit cells
in the $ac$ plane.}
\end{figure}

Fig.~\ref{fig2} displays the X-ray powder diffraction data
obtained from a crushed piece of the \bmo\ single crystal. The
structure refinement with these data using
Fullprof2k\cite{roisnel01} yields the unit-cell parameters $a =
4.3819(1)$~\AA, $b = 10.9745(5)$~\AA, and $c = 3.55329(8)$~\AA, in
good agreement with $a = 4.385$~\AA, $b = 10.967$~\AA, and $c =
3.552$~\AA\ reported previously\cite{sander79}. Our
room-temperature X-ray diffraction data perfectly agree with the
already reported $Immm$ symmetry. However, the low-temperature
data (at 150 and 100~K) of a \bmo\ single crystal, see
Table~\ref{tab1}, clearly indicate a lower monoclinic symmetry
$C2/m$ (see inset in Fig.~\ref{fig2}), as choices with higher
symmetries cannot index all observed intensities.

\begin{figure}
\includegraphics[width= 0.8\linewidth]{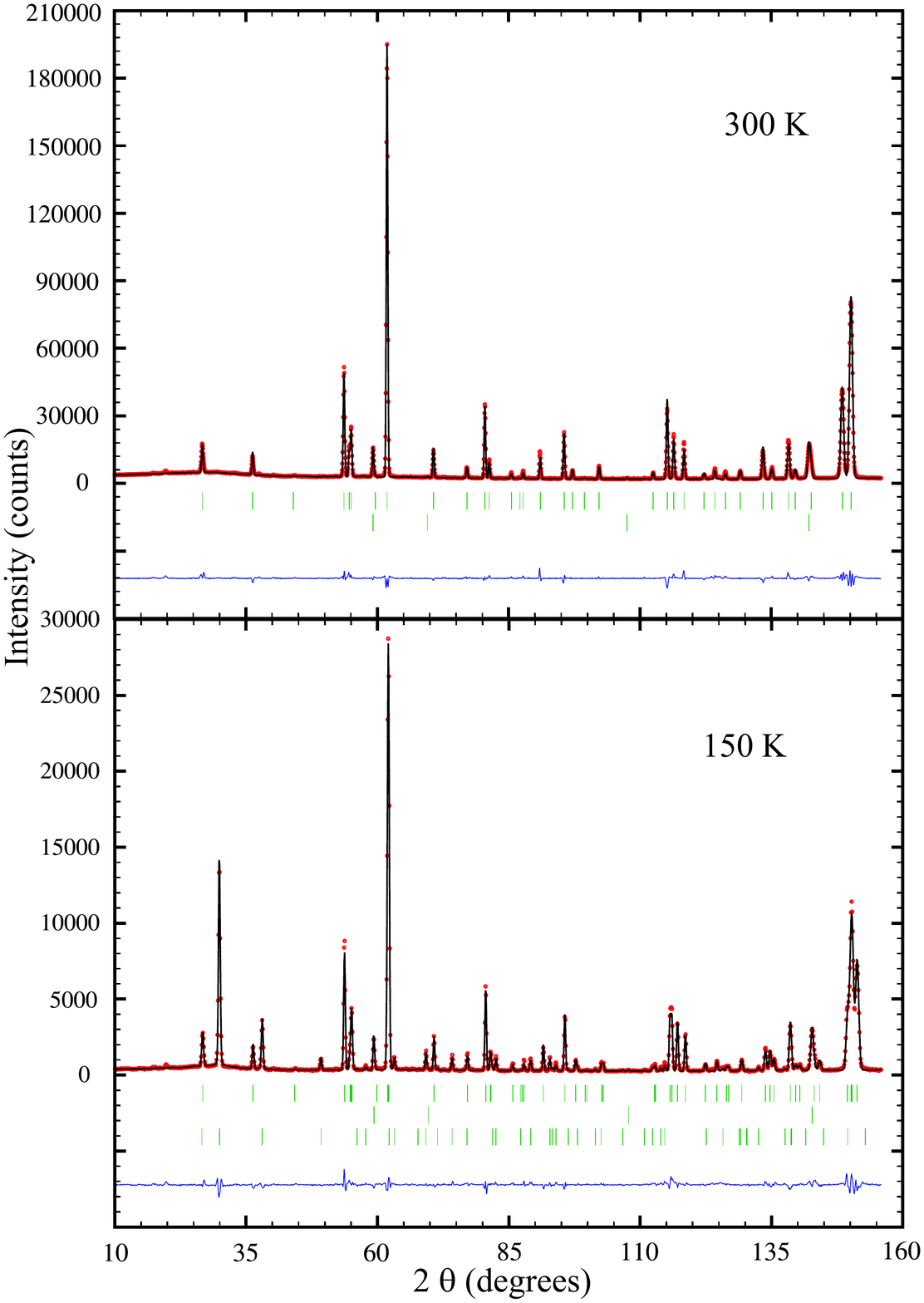}
\caption{\label{fig5} (Color online) Neutron diffraction data of \bmo\ at 300 (top) and
150~K (bottom). Observed intensities are marked by circles and the
calculated pattern as a line. All Bragg intensities are indicated
by vertical lines: \bmo\ (1st row), MnO (2nd row), and \bmo\
spin structure (3rd row). The lowest line is the difference between observed
and calculated intensities $I_{obs}-I_{calc}$.}
\end{figure}

In Fig.~\ref{fig5}, two representative neutron powder diffraction
patterns are shown and the results of Rietveld refinements for
five different temperatures are summarized in Table~\ref{tab2}.
The data of Fig.~\ref{fig5} reveal that the sample contains about
3~\% of MnO as an impurity phase. This impurity phase in the
neutron data probably originates from the fact that the powdered
sample was exposed to air before placing it in the vanadium cup. A
partial decomposition of the sample explains this
minor complication in the neutron data, but there is no evidence
of an antiferromagnetic ordering transition at 118~K in the
magnetic investigations, which would be expected if a significant
amount of MnO was present in the original single crystal as well.

\begin{table*}
\caption{Atomic and magnetic parameters of \bmo\ at different temperatures. 
All atomic positions are assumed
to be fully occupied and STDs are already multiplied by the 
Berar-Lelann factor\cite{berar91} and an additional factor of three to compensate for pronounced texture effects in the raw data~\cite{texture}. Data for MnO was taken
from G.R. Levi\cite{levi24} and Goodwin {\it et al.}\cite{goodwin06}.}
 \label{tab2}

\vskip1mm
\parbox{17.8cm}{
\begin{ruledtabular}
\begin{tabular}{llllll}
Temperature (K) & 300  & 200  & 150  & 100  & 3  \\ \hline 
Ba $(x, y, z, U_{iso})$	& 1/2, 1/2, 0, 0.0073(5)& 1/2, 1/2, 0, 0.0045(4) &	0, 0, 0, 0.0054(7)&	0, 0, 0, 0.0073(6) & 0, 0, 0, 0.0033(9) \\ \hline 
Mn $(x, y, z, U_{iso})$	& 1/2, 0.1882(1), 0,	& 1/2, 0.18811(9), 0,	& 0, 0.3125(9), 0, & 0, 0.3124(9), 0,  &	0, 0.312(1), 0, \\
 & 0.0130(5)	& 0.0132(4)	&  0.0085(7)& 0.0094(7) &	 0.0075(9) \\ \hline 
O1 $(x, y, z, U_{iso})$	& 0, 1/2, 1/2, 0.0190(5) &	0, 1/2, 1/2, 0.0172(4)	& 0, 1/2, 0, 0.0116(7)	& 0, 1/2, 0, 0.0127(7) &	0, 1/2, 0, 0.0144(9)\\  \hline 
O2 $(x, y, z, U_{iso})$	& 0, 0.2048(1), 0, & 0, 0.20423(9), 0,	& 0, 0.2039(2), 1/2, 	& 0, 0.20402(9), 1/2,  &	0, 0.2030(1), 1/2, \\  
	&  0.0176(4)	&  0.0161(3)	&  0.0157(5)	&  0.0154(5) & 0.0101(6)\\ \hline 
Space Group (Nr.)& 	$Immm$ (71)& $Immm$ (71)	& $C2/m$ (12)	&$C2/m$ (12)	&$C2/m$ (12)	\\ \hline 
$a$~(\AA) &	4.3834(1)&	4.38006(6)	&5.62060(6)	& 5.61194(6)	& 5.60595(6) \\ \hline 
$b$~(\AA) &	10.9751(2)&	10.96636(9)	&10.96155(9)&	10.9577(1)	&10.9553(1)\\ \hline 
$c$~(\AA) &	3.5549(1)	& 3.54433(6)	& 3.53794(6)	& 3.53324(6)	& 3.53000(6)\\  \hline 
$\beta$~($^\circ$) &	90	& 90	& 128.8199(3)	& 128.7261(3)& 	128.6678(3)  \\ \hline 
$\chi^2$ 	& 4.23	& 2.23	& 7.92	& 4.35	& 5.38\\  \hline 
R$_{\rm Bragg}$ &	0.0572	& 0.0448	& 0.0482	& 0.0584	& 0.0704\\ \hline 
R$_{\rm F}$ &	0.0556&	0.0412&	0.0443&	0.0433&	0.0529\\ \hline 
Magn. $k$ vector	& ---	& ---	& [0,0,1/2]& 	[0,0,1/2]&	[0,0,1/2]\\ \hline 
R$_{\rm mag}$ & ---	& ---	& 0.117& 	0.0985& 	0.0955\\  \hline 
Comment& 	MnO 3.13(6)\%& 	MnO 3.09(6)\%  & 	MnO 2.94(6)\% &	MnO excluded & MnO	excluded 
\end{tabular}
\end{ruledtabular}
} 
\end{table*}

\begin{figure}
\includegraphics[width= 0.8\linewidth]{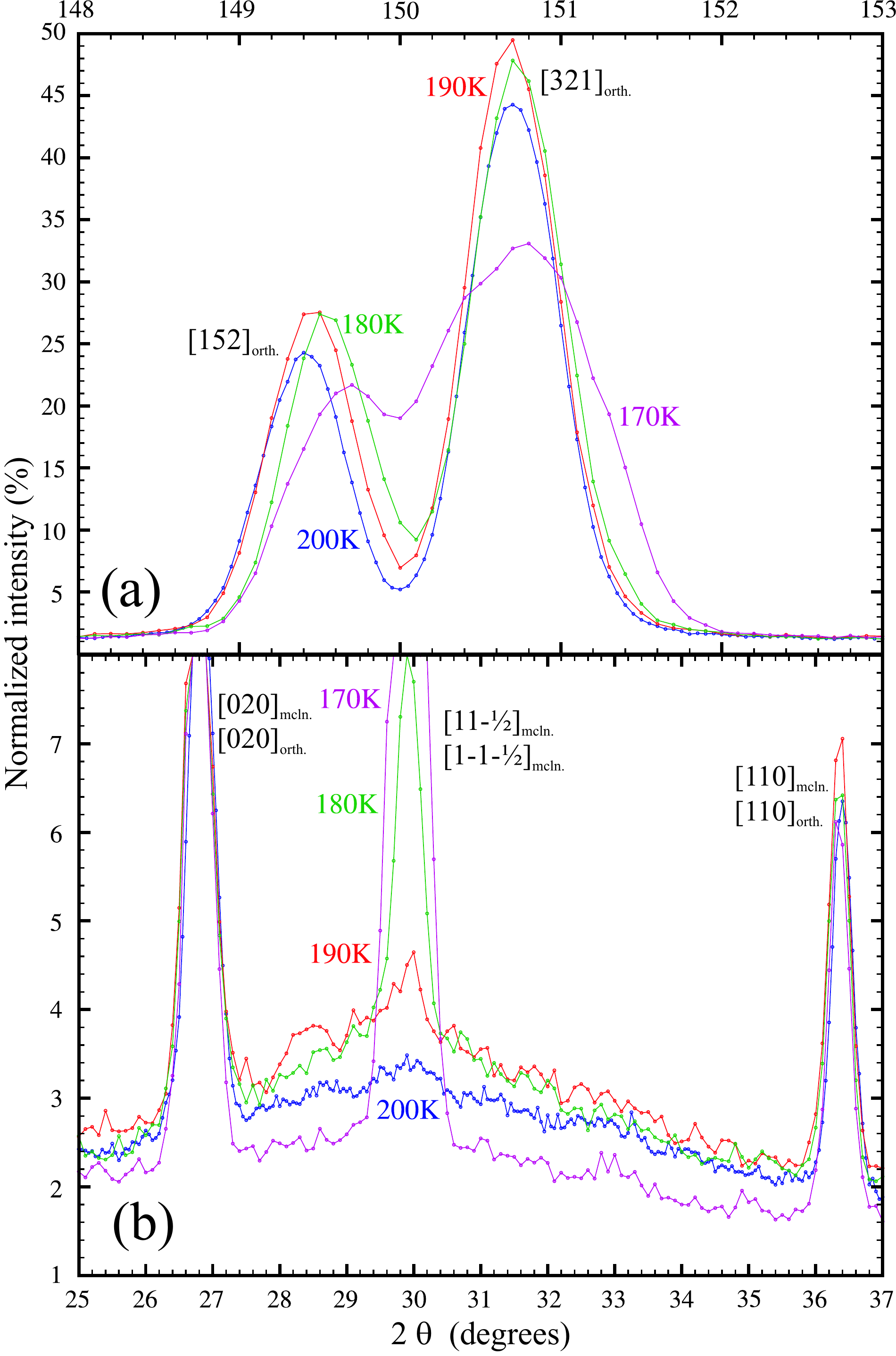}
\caption{\label{fig6} (Color online) Representative neutron diffraction data at
four temperatures. (a) Two of the peaks from the high-temperature
orthorhombic structure are indexed and the splitting of,
especially the [321] peak, is made clear. (b) The growth of the
largest magnetic peak displayed at the same temperatures as in (a)
and indexed according to high- and low-temperature
crystallographic settings.}
\end{figure}

According to the thermodynamic data (see above), \bmo\ orders
magnetically at 184~K and, indeed, magnetic Bragg peaks are
visible at 180~K. Simultaneously, several nuclear peaks split
indicating that the symmetry is lowered, in agreement with the
lowering to the monoclinic symmetry obtained from the
low-temperature single crystal X-ray data. An example of the
splitting is shown in Fig.~\ref{fig6}a; the two intensities,
indexed as [152] ($2\Theta = 149.5^\circ$) and [321] ($2\Theta =
150.6^\circ$) in the centered orthorhombic setting ($Immm$), shift
only slightly between 200 and 190~K, due to shrinkage of the unit
cell, but at 170~K both peaks are clearly split and their
intensities are distributed. Due to the high scattering angle we
may safely neglect any magnetic scattering as possible cause of
the peak splitting.

The emergence of the magnetic Bragg peaks upon cooling is shown
for the example of the (1 1 -1/2)/(1 -1 -1/2) reflection in Fig.
6b). This purely magnetic intensity develops from a diffuse
scattering, which is present already at 200~K as a broad hump on
top of the background intensity. At 190 K, the diffuse scattering
sharpens, which means that the spin-spin correlation length
increases, and at 180~K, this peak exhibits a well-defined half
width, close to that of the nuclear scattering intensities at
similar scattering angle, e.g.\ [020]$_{mcln.}$

\begin{figure}
\includegraphics[width= 0.8\linewidth]{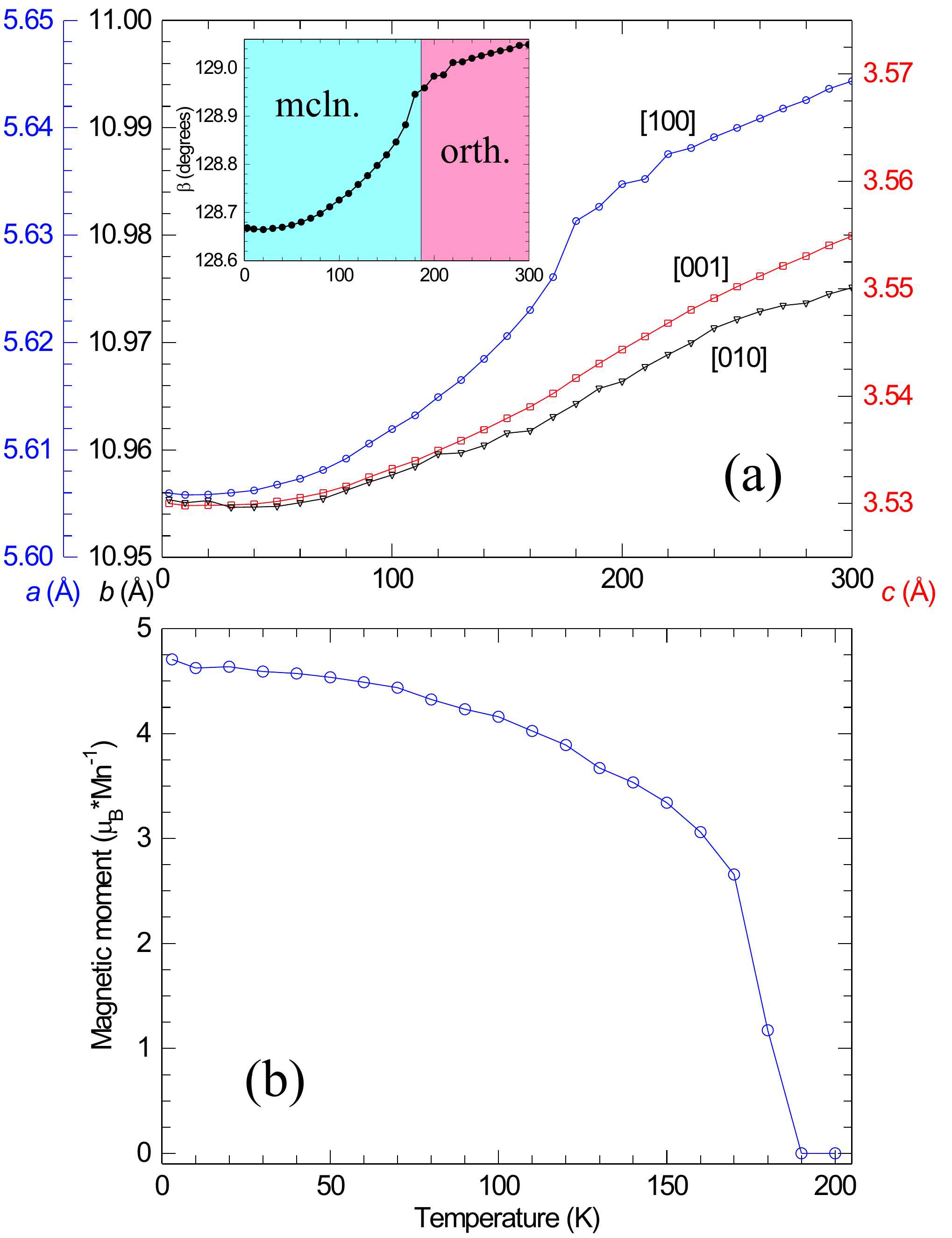}
\caption{\label{fig7} (Color online) (a) The monoclinic unit-cell parameters, the $\beta$ angle (inset), 
and (b) the ordered magnetic moment of Mn$^{2+}$ as functions of temperature.}
\end{figure}

To enable unit cell comparisons between all temperatures, the
monoclinic unit cell was refined on all diffraction data and the
result is shown in Fig.~\ref{fig7}. The lattice constants $a$,
$b$, and $c$ decrease continuously on cooling between 300 and
50~K, but $a$ changes much stronger than $b$ and $c$. A small
anomaly in the $a$ parameter is visible on cooling through $T_N$.
As shown in the inset of Fig.~\ref{fig7}, the monoclinic angle
$\beta$ indicates that the structure distorts below 180 K. Above
200~K, the value of $\beta$ was fixed to the value calculated via
the orthorhombic lattice constants, see inset of Fig.~\ref{fig2}.
In the bottom panel of Fig.~\ref{fig7} we show the ordered
Mn$^{2+}$ magnetic moment as a function of temperature.

\begin{figure}
\includegraphics[width= 0.99\linewidth]{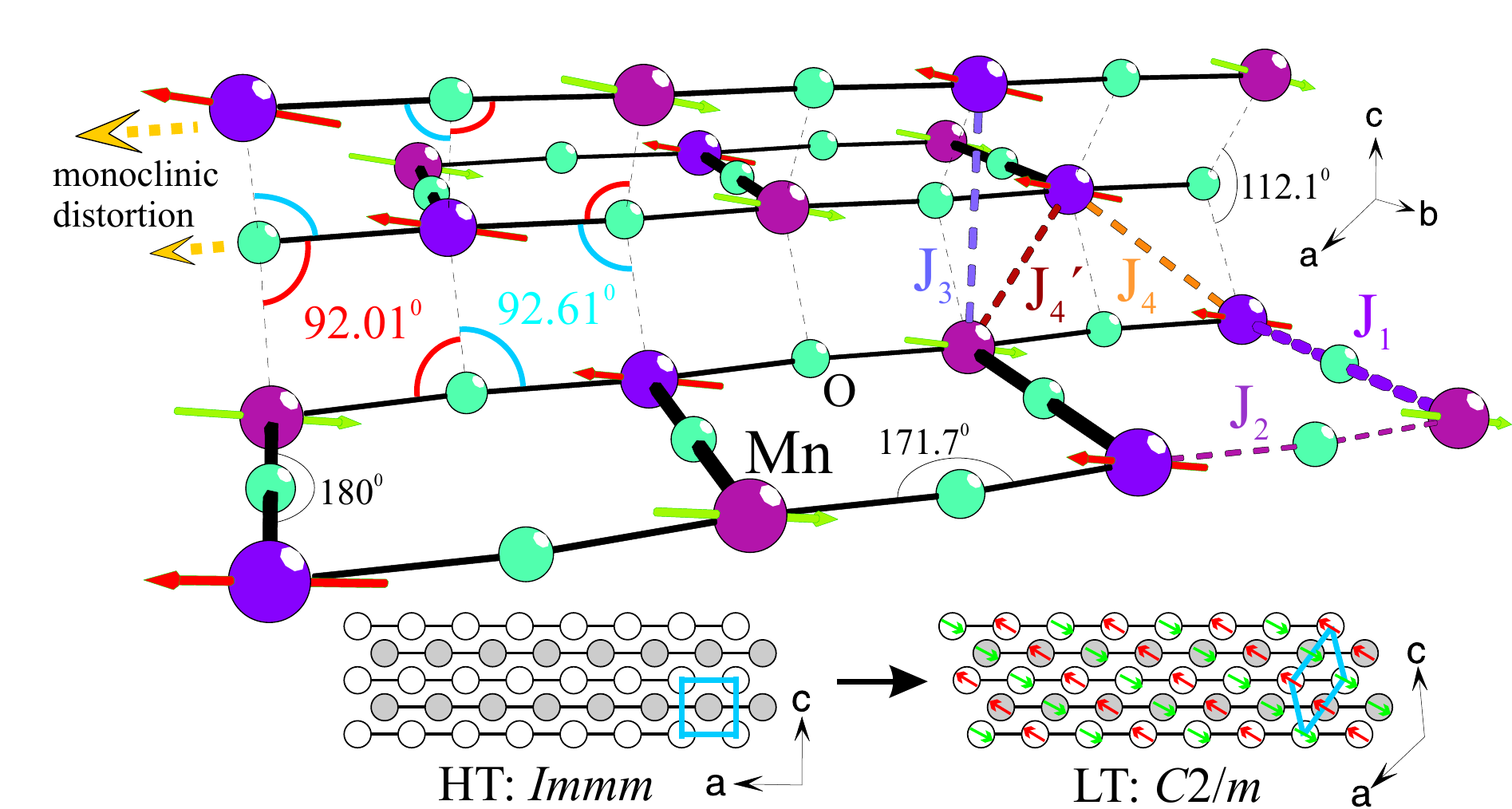}
\caption{\label{fig9} (Color online) The magnetic structure of \bmo\ at 3 K,
represented on its Mn-O sublattice. The magnetic moments of Mn are
represented by arrows. The thick dashed lines represent the 
different Mn--to--Mn couplings ($J_1$--$J_4$).
The arrows on the left show the ladder shift through
the monoclinic distortion, as discussed in the text. The bottom panels 
compare the high-temperature (HT) and low-temperature structures of \bmo.}
\end{figure}

All magnetic Bragg peaks appearing below $T_N\simeq 184$~K can be
indexed with the propagation vector $k$=(0 0 1/2), i.e. by a
doubling of the $c$ axis. This means that the next-nearest
neighbor ladders connected by one $c$-lattice spacing are ordering
antiferromagnetically, whereas the next-nearest neighbor ladders
connected by one $b$-lattice spacing are coupled
ferromagnetically. The coupling along $c$ is mediated by the $J_3$
parameter, which the magnetic symmetry determines to be
antiferromagnetic. Due to the 180$^\circ$ Mn-O-Mn bonds there is
no doubt that the intra-ladder interaction parameters $J_1$ and
$J_2$ must be antiferromagnetic, which unambiguously fixes the
magnetic order within a single ladder. With these arguments we
determine the magnetic structure including the inter-ladder
coupling between ladders connected by one $c$ or by one $a$
lattice parameter. What has not been discussed so far is the
coupling between nearest-neighbor ladders connected by the body
centering of the orthorhombic lattice (1/2,1/2,1/2). However, the
two choices of ferromagnetic and antiferromagnetic coupling do not
result in different symmetries but only in two different domain types. 
Choosing a ferromagnetic coupling for (1/2,1/2,1/2) yields
an antiferromagnetic coupling for (1/2,-1/2,1/2) and {\it vice versa}.
The magnetic symmetry constructed above is monoclinic and the two
domain types correspond to the common symmetry reduction from
orthorhombic to monoclinic. As already discussed above, the Mn
moment is dominantly oriented along the $a$ axis and a picture of
the spin structure in comparison with the ladders is shown in
Fig.~\ref{fig9}. With this magnetic model the neutron
powder-diffraction data is perfectly described. The Mn$^{2+}$
ordered moment rapidly increases and half of the expected
$5~\mu_B$ is observed already at 170~K. With further decreasing
temperature, the moment further increases and finally saturates at
$\simeq 4.7~\mu_B$ close to the expected value.

The role of the monoclinic structural distortion in \bmo \ has to
be discussed concerning two aspects. First, the monoclinic
distortion is just a consequence of the monoclinic magnetic
symmetry; but thereby one would only expect a weak structural
deformation. Second, the monoclinic distortion lifts the magnetic
frustration and thereby stabilizes the magnetic order in \bmo \
significantly. The nearest-neighbor inter-ladder interaction
parameters $J_4$ and $J_4'$  are mediated via Mn--O--Mn bonds
close to a 90$^\circ$ configuration. In the high-temperature
phase, both Mn--O--Mn angles are $92.3^\circ$ (Fig.~\ref{fig1}b),
which causes a perfect geometrical frustration. On cooling through
$T_N$, the structure changes from orthorhombic to monoclinic and
induces an antisymmetric shift of the ladders, which changes the
corresponding  Mn--O--Mn angles by $\pm 0.3^\circ$ and lifts the 
perfect geometrical frustration. According to the Goodenough-Kanamori 
rules\cite{goodenough55,kanamori59}, a decreasing Mn--O--Mn angle results 
in a monotonic decrease from a large antiferromagnetic coupling at an  
180$^\circ$ bond to a weak ferromagnetic coupling at a
90$^\circ$ bond. Thus one might expect $J_4'$ with the larger angle of
$92.61^\circ$ to be increased while $J_4$ with the smaller
angle of $92.01^\circ$ to be decreased. This is, however, in contrast 
to our neutron data, which clearly show that the neighboring spins coupled via 
the larger bond angle of $92.61^\circ$ are aligned parallel; see Fig.~\ref{fig9}.   
This means that $J_4$ is larger than $J_4'$ what most probably means that the 
minimum in the Mn--O--Mn angular dependence is located at a bond angle above 
93$^\circ$.

Due to the strong sensitivity of the
magnetic interaction on this bond angle a sizeable magnetoelastic
effect occurs in \bmo , similar to the spin-Peierls compound
CuGeO$_3$ where dimerization mainly arises from a Cu-O-Cu bond-angle modulation\cite{geertsma,braden}. A similar lifting of
geometric frustration through a coupled structural and magnetic
transition has been reported recently for VOCl\cite{komarek} and
seems to be relevant in the parent phase of the iron-arsenide
superconductors as well. In LaOFeAs the magnetic order is
associated with a tetragonal-to-orthorhombic transition which also
has a two-fold role\cite{laofeas}. It is a consequence of the
lower magnetic symmetry and it lifts a magnetic frustration thereby
stabilizing the magnetic order in full analogy to the discussion
presented above for \bmo .

\section{Conclusions}

We have presented  a detailed study of large single crystals of
\bmo\ grown in a floating-zone image furnace. \bmo\ has structural
motifs from both the hexagonal $R$MnO$_3$\cite{zhou06} and the
well-known spin-ladder compound
(Sr,Ca)$_{14}$Cu$_{24}$O$_{41}$\cite{kazakov}. Although \bmo\
contains spin ladders with obvious geometrical frustration, this
compound shows long-range antiferromagnetic order at $T_N\simeq
184$~K. The magnetic ordering is accompanied by a distortion of
the crystal structure, which is described by a lowering of the
symmetry from orthorhombic to monoclinic. Data from both
specific-heat and thermal-expansion measurements clearly indicate
that this phase transition shows a weak temperature hysteresis.
The weak first-order feature of this magnetically driven phase
transition is attributed to a strong magnetoelastic
coupling. The magnetic correlations above $T_N$ are most likely
confined within the ladders, which cannot show long-range order
due to the low dimensionality of the (almost) isotropic Heisenberg
spin system. The low dimensionality of the magnetic subsystem of
\bmo\ is a consequence of a  very effective magnetic decoupling
between neighboring spin ladders due a perfect geometrical
frustration. Experimentally, this low dimensionality is reflected
by broad maxima in the temperature dependence of the magnetic
suceptibility, which are located around 270~K suggesting an
average intra-ladder coupling $J\simeq 40$~K. The low
dimensionality is also clearly seen in the neutron diffraction
data, where pronounced short-range magnetic correlations are
observed already well above $T_N$. The neutron data reveal that at
$T_N$ the magnetic correlations become long-range and, in
addition, there is an antisymmetric shift of neighboring spin
ladders, which results in an alternation of the corresponding
Mn--O--Mn angles along the ladder direction. As a consequence, the
geometrical frustration is lifted, the spin ladders become
three-dimensionally coupled and, simultaneously, the magnetic
subsystem develops long-range antiferromagnetic order. This
indicates, that the structural phase transition from orthorhombic
to monoclinic in \bmo\ is driven by a gain in the magnetic free
energy.

\acknowledgements

We would like to thank Daniel Khomskii for his invaluable comments
and Inge Simons for performing the elemental analyses. This work
was supported by the Deutsche Forschungsgemeinschaft through
Sonderforschungsbereich~608.


\begin{thebibliography}{10}
\parskip-0.2ex plus0.05ex minus0.05ex


\bibitem{dagotto92}
E.~Dagotto, J.~Riera, and D.~Scalapino.
\newblock Phys.\ Rev.\ B {\bf 45}, 5744 (1992).

\bibitem{white02}
S.R. White, I.~Affleck, and D.J. Scalapino.
\newblock Phys.\ Rev.\ B {\bf 65}, 165122 (2002).

\bibitem{carr02}
S.T. Carr and A.M. Tsvelik.
\newblock Phys.\ Rev.\ B {\bf 65}, 195121 (2002).

\bibitem{vuletic06}
T.~Vuletic, B.~Korin-Hamzic, T.~Ivek, S.~Tomic, B.~Gorshunov, M.~Dressel, and
  J.~Akimitsu.
\newblock {\bf 428}, 169 (2006).

\bibitem{hiroi91}
Z.~Hiroi, M.~Azuma, M.~Takano, and Y.~Bando.
\newblock J.\ of Solid State Chem. {\bf 95}, 230 (1991).

\bibitem{dagotto96}
E.~Dagotto and T.M. Rice.
\newblock Science {\bf 271}, 618 (1996).

\bibitem{goodenough55}
J.B. Goodenough.
\newblock Phys.\ Rev. {\bf 100}, 564 (1955).

\bibitem{kanamori59}
J.~Kanamori.
\newblock J.\ Phys.\ Chem.\ Solids {\bf 10}, 87 (1959).

\bibitem{eccleston96}
R.S. Eccleston, M.~Azuma, and M.~Takano.
\newblock Phys.\ Rev.\ B {\bf 53}, 14721 (1996).

\bibitem{martin-delgado98}
M.~Martin-Delgado, J.~Dukelsky, and G.~Sierra.
\newblock Physics Lett.\ A {\bf 250}, 430 (1998).

\bibitem{Windt01}
M.~Windt, M.~Gr\"uninger, T.~Nunner, C.~Knetter, K.P.~Schmidt,
 G.S.~Uhrig, T.~Kopp, A.~Freimuth,
     U.~Ammerahl, B.~B\"uchner, and A.~Revcolevschi,
  Phys. Rev. Lett. {\bf 87}, 127002 (2001).

\bibitem{Klanjsek08}
M. Klanj\v{s}ek, H. Mayaffre, C. Berthier, M. Horvati\'c, B.
Chiari, O. Piovesana, P. Bouillot, C. Kollath, E. Orignac, R.
Citro, and T. Giamarchi,
  Phys. Rev. Lett. {\bf 101},  137207  (2008).

\bibitem{Ruegg08cm}
Ch. R\"uegg, K. Kiefer, B. Thielemann, D.~F. McMorrow, V. Zapf, B.
Normand, M.~B. Zvonarev, P. Bouillot, C. Kollath, T. Giamarchi, S.
Capponi, D. Poilblanc, D. Biner, and K. Kr\"amer,
 Phys. Rev. Lett. {\bf 101}, 247202 (2008) .

\bibitem{Thielemann09PRL}
B. Thielemann, Ch. R\"uegg, H. M. R{\o}nnow, A. M. L\"auchli, J.-S.
Caux, B. Normand, D. Biner, K.W. Kr\"amer, H.-U. G\"udel, J.
Stahn, K. Habicht, K. Kiefer, M. Boehm, D. F. McMorrow, and J.
Mesot,
 Phys. Rev. Lett. {\bf 102},  107204  (2009).

\bibitem{Sologubenko00}
A.~V. Sologubenko, K. Giann\`o, H.~R. Ott, U. Ammerahl, and A.
Revcolevschi,
  Phys. Rev. Lett. {\bf 84},  2714  (2000).

\bibitem{Hess01}
C. Hess, C. Baumann, U. Ammerahl, B. B\"uchner, F.
Heidrich-Meisner, W. Brenig, and A. Revcolevschi,
  Phys. Rev. B {\bf 64},  184305  (2001).

\bibitem{Lorenz08}
T. Lorenz, O. Heyer, M. Garst, F. Anfuso, A. Rosch, C. R\"uegg,
and K. Kr\"amer,
 Phys. Rev. Lett. {\bf 100},  067208  (2008).

\bibitem{Anfuso08}
F. Anfuso, M. Garst, A. Rosch, O. Heyer, T. Lorenz, C. R\"uegg,
and K. Kr\"amer,
 Phys. Rev. B {\bf 77},  235113  (2008).

\bibitem{sander79}
K.~Sander and H.~K. M\"{u}ller-Buschbaum.
\newblock Z.\ Anorg.\ Allg.\ Chem. {\bf 451}, 35 (1979).

\bibitem{tangoulis07}
V.~Tangoulis.
\newblock Chem.\ Phys. {\bf 332}, 271 (2007).

\bibitem{zhou06}
J.S. Zhou, J.B. Goodenough, J.M. Gallardo-Amores, E.~Moran, M.A. Alario-Franco,
  and R.~Caudillo.
\newblock Phys.\ Rev.\ B {\bf 74}, 014422 (2006).

\bibitem{kihlborg68}
L.~Kihlborg and W.~Israelss.m.
\newblock Acta Chem. Scand. {\bf 22}, 1685 (1968).

\bibitem{lashley03}
J.C. Lashley, M.F. Hundley, A.~Migliori, J.L. Sarrao, P.G. Pagliuso, T.W.
  Darling, M.~Jaime, J.C. Cooley, W.L. Hults, L.~Morales, D.J. Thoma, J.L.
  Smith, J.~Boerio-Goates, B.F. Woodfield, G.R. Stewart, R.A. Fisher, and N.E.
  Phillips.
\newblock Cryogenics {\bf 43}, 369 (2003).

\bibitem{dejongh74}
L.J. Dejongh and A.R. Miedema.
\newblock Adv.\ in Phys. {\bf 23}, 1 (1974).

\bibitem{kadanoff67}
L.P. Kadanoff, W.~Gotze, D.~Hamblen, R.~Hecht, E.A.S. Lewis, W.~Palciaus.vv,
  M.~Rayl, J.~Swift, D.~Aspnes, and J.~Kane.
\newblock Rev.\ Mod.\ Phys. {\bf 39}, 395 (1967).

\bibitem{ito78}
T.~Ito, K.~Ito, and M.~Oka.
\newblock Jpn.\ J.\ Appl.\ Phys. {\bf 17}, 371 (1978).

\bibitem{roisnel01}
T.~Roisnel and J.~Rodriguez-Carvajal.
\newblock Fullprof2k V. 1.8a Laboratoire Leon Brillouin (CEA-CNRS) 91191
  Gif-sur-Yvette Cedex (France) (2001).

\bibitem{berar91}
J.F. Berar and P.~Lelann.
\newblock J. Appl. Crystallogr. {\bf 24}, 1 (1991).

\bibitem{texture}
The full two-dimensional analysis of the neutron powder data exhibited internal structure, because the powder contained comparatively large grains. To correct this effect in the Rietveld refinement we used an enhanced absorption correction and multiplied all standard deviations from the  refinements by an additional factor of three. This factor was estimated by comparing the data from the refinement with the larger absorption with those from a refinement with a correct absorption correction which, however, yields negative thermal displacements.

\bibitem{levi24}
G.~R. Levi.
\newblock Rendiconti dell'Istituto Lombardo di Science e Lettere Classe di
  Science Mathematiche e Naturali {\bf 57}, 619 (1924).

\bibitem{goodwin06}
A.L. Goodwin, M.G. Tucker, M.T. Dove, and D.A. Keen.
\newblock Phys.\ Rev.\ Lett. {\bf 96}, 047209 (2006).

\bibitem{kazakov}
S.M. Kazakov, J.~Karpinski, G.I. Meijer, C.~Bougerol-Chaillout, and
  M.~Nunez-Regueiro.
\newblock Physica C {\bf 351}, 301 (2001).

\bibitem{geertsma} W. Geertsma and D. Khomskii.
Phys. Rev. B {\bf 54}, 3011 (1996).

\bibitem{braden} M. Braden, G. Wilkendorf,
J. Lorenzana, M. A\"in, G.J. McIntyre, M. Behruzi, G. Heger, G.
Dhalenne, and A. Revcolevschi, Phys. Rev. B {\bf 54}, 1105
(1996).

\bibitem{komarek}A.C. Komarek, T. Taetz, M.T. Fern\'andez-D\'iaz, D.M. Trots, A. M\"oller, and M.
Braden. Phys. Rev. B {\bf 79}, 104425 (2009).

\bibitem{laofeas}T. Yildirim. Phys. Rev. Lett. {\textbf 101}, 057010, (2008).

\end{thebibliography}
\end{document}